\documentclass[conference, a4paper]{IEEEtran}
\IEEEsettopmargin{t}{1.92cm}

\IEEEoverridecommandlockouts
\usepackage{cite}
\usepackage{amsmath,amssymb,amsfonts}
\usepackage{algorithm}
\usepackage{algpseudocode}
\usepackage{graphicx}
\usepackage{textcomp}
\usepackage{xcolor}
\usepackage{todonotes}
\makeatletter
\let\MYcaption\@makecaption
\makeatother
\usepackage[font=footnotesize,labelformat=simple]{subcaption}

\makeatletter
\let\@makecaption\MYcaption
\makeatother
\usepackage{tikz}
\usepackage{comment}
\usepackage{pgfplots}
\usetikzlibrary{shapes,arrows,positioning,calc, matrix,spy,patterns,backgrounds,shapes.geometric}

\usepackage[nolist]{acronym}
\newcommand*\dif{\mathop{}\!\mathrm{d}}
\pgfplotsset{compat=1.18}

\usepackage{booktabs}
\usepackage{xfrac}
\usepackage{comment}

\definecolor{cb-1}{HTML}{4477AA}
\definecolor{cb-2}{HTML}{EE6677}
\definecolor{cb-3}{HTML}{228833}
\definecolor{cb-4}{HTML}{CCBB44}
\definecolor{cb-5}{HTML}{66CCEE}
\definecolor{cb-6}{HTML}{AA3377}
\definecolor{cb-7}{HTML}{BBBBBB}

\pgfplotscreateplotcyclelist{cb list}{
	cb-1,cb-2,cb-3,cb-4,cb-5,cb-6,cb-7
}

\definecolor{kit-green100}{rgb}{0,.59,.51}
\definecolor{kit-green70}{rgb}{.3,.71,.65}
\definecolor{kit-green50}{rgb}{.50,.79,.75}
\definecolor{kit-green30}{rgb}{.69,.87,.85}
\definecolor{kit-green15}{rgb}{.85,.93,.93}
\definecolor{KITgreen}{rgb}{0,.59,.51}

\definecolor{KITpalegreen}{RGB}{130,190,60}
\colorlet{kit-maigreen100}{KITpalegreen}
\colorlet{kit-maigreen70}{KITpalegreen!70}
\colorlet{kit-maigreen50}{KITpalegreen!50}
\colorlet{kit-maigreen30}{KITpalegreen!30}
\colorlet{kit-maigreen15}{KITpalegreen!15}

\definecolor{KITblue}{rgb}{.27,.39,.66}
\definecolor{kit-blue100}{rgb}{.27,.39,.67}
\definecolor{kit-blue70}{rgb}{.49,.57,.76}
\definecolor{kit-blue50}{rgb}{.64,.69,.83}
\definecolor{kit-blue30}{rgb}{.78,.82,.9}
\definecolor{kit-blue15}{rgb}{.89,.91,.95}

\definecolor{KITyellow}{rgb}{.98,.89,0}
\definecolor{kit-yellow100}{cmyk}{0,.05,1,0}
\definecolor{kit-yellow70}{cmyk}{0,.035,.7,0}
\definecolor{kit-yellow50}{cmyk}{0,.025,.5,0}
\definecolor{kit-yellow30}{cmyk}{0,.015,.3,0}
\definecolor{kit-yellow15}{cmyk}{0,.0075,.15,0}

\definecolor{KITorange}{rgb}{.87,.60,.10}
\definecolor{kit-orange100}{cmyk}{0,.45,1,0}
\definecolor{kit-orange70}{cmyk}{0,.315,.7,0}
\definecolor{kit-orange50}{cmyk}{0,.225,.5,0}
\definecolor{kit-orange30}{cmyk}{0,.135,.3,0}
\definecolor{kit-orange15}{cmyk}{0,.0675,.15,0}

\definecolor{KITred}{rgb}{.63,.13,.13}
\definecolor{kit-red100}{cmyk}{.25,1,1,0}
\definecolor{kit-red70}{cmyk}{.175,.7,.7,0}
\definecolor{kit-red50}{cmyk}{.125,.5,.5,0}
\definecolor{kit-red30}{cmyk}{.075,.3,.3,0}
\definecolor{kit-red15}{cmyk}{.0375,.15,.15,0}

\definecolor{KITpurple}{RGB}{160,0,120}
\colorlet{kit-purple100}{KITpurple}
\colorlet{kit-purple70}{KITpurple!70}
\colorlet{kit-purple50}{KITpurple!50}
\colorlet{kit-purple30}{KITpurple!30}
\colorlet{kit-purple15}{KITpurple!15}

\definecolor{KITcyanblue}{RGB}{80,170,230}
\colorlet{kit-cyanblue100}{KITcyanblue}
\colorlet{kit-cyanblue70}{KITcyanblue!70}
\colorlet{kit-cyanblue50}{KITcyanblue!50}
\colorlet{kit-cyanblue30}{KITcyanblue!30}
\colorlet{kit-cyanblue15}{KITcyanblue!15}

\definecolor{cb-1}{HTML}{4477AA}
\definecolor{cb-2}{HTML}{EE6677}
\definecolor{cb-3}{HTML}{228833}
\definecolor{cb-4}{HTML}{CCBB44}
\definecolor{cb-5}{HTML}{66CCEE}
\definecolor{cb-6}{HTML}{AA3377}
\definecolor{cb-7}{HTML}{BBBBBB}

\pgfplotscreateplotcyclelist{cb list}{
	cb-1,cb-2,cb-3,cb-4,cb-5,cb-6,cb-7
}

\let\j\relax
\newcommand{\j}{\mathrm{j}}

\newcommand*{\vect}[1]{\boldsymbol{#1}}
\newcommand*{\mat}[1]{\MakeUppercase{\boldsymbol{#1}}}

\usepackage{ellipsis}
\usetikzlibrary{calc, patterns,fit,shapes,positioning}
\usetikzlibrary{decorations.pathreplacing,decorations.markings,shapes.geometric}
\tikzset{naming/.style={align=center,font=\small}}
\tikzset{antenna/.style={insert path={-- coordinate (ant#1) ++(0,0.5) edge[-] +(135:0.5) + (0,0) edge[-] +(45:0.5)}}}
\tikzset{station/.style={naming,draw,shape=dart,shape border rotate=90}}
\tikzset{mobile/.style={naming,draw,shape=rectangle,minimum width=12mm,minimum height=6mm, outer sep=0pt,inner sep=3pt}}
\tikzset{radiation/.style={{decorate,decoration={expanding waves,angle=90,segment length=4pt}}}}
\usetikzlibrary{decorations.pathreplacing}
\usetikzlibrary{arrows,calc,fit,matrix,positioning,shapes,shadows,trees,mindmap,tikzmark,arrows.meta,angles,quotes,babel,spy}

\tikzset{pics/MUE/.style args={#1}
{code={
\node[mobile,label={[inner ysep=+.3333em]\dots}] (box){};
\draw ([xshift=.25cm] box.south west) circle (4pt)
      ([xshift=-.25cm]box.south east) circle (4pt);

\fill ([xshift=.25cm] box.south west) circle (1pt)
      ([xshift=-.25cm]box.south east) circle (1pt);

\draw ([xshift=.25cm] box.north west) [antenna=1];
\draw ([xshift=-.25cm]box.north east) [antenna=2];
}}}

\tikzset{pics/UE/.style args={#1}
{code={
\node[mobile] (box) {#1};

\draw ([xshift=.25cm] box.south west) circle (4pt)
      ([xshift=-.25cm]box.south east) circle (4pt);

\fill ([xshift=.25cm] box.south west) circle (1pt)
      ([xshift=-.25cm]box.south east) circle (1pt);

\draw (box.north) [antenna=1];
}}}

\tikzset{pics/MBS/.style args={#1}{code={
\node[coordinate] (-base) at (0,0)  {};
\node[coordinate] (-l) at (-2,-1) {};
\node[coordinate] (-r) at (2,-1) {};
\node[coordinate] (-t) at (0,4) {};
\draw[line join=bevel] (-base) -- (-l) -- (-t) -- (-r) -- cycle;
\draw[line join=bevel] ($(-l)!0.5!(-t)$) -- ($(-r)!0.5!(-t)$) -- ($(-l)!0.7!(-t)$) -- ($(-r)!0.7!(-t)$) -- cycle;
\draw[line join=bevel] ($(-l)!0.7!(-t)$) -- ($(-r)!0.8!(-t)$) -- ($(-l)!0.8!(-t)$) -- ($(-r)!0.7!(-t)$) -- cycle;
\node[coordinate, label={[label distance=0.5]above:\dots}] (-d) at (0,4) {};
\draw[line cap=rect,-] ($ (-t)+(-1,0)$) [antenna=1];
\draw[line cap=rect,-] ($ (-t)+(-1,0)$) -- ($(-t)+(1,0)$) [antenna=2];
\node[coordinate] (-a1) at ($ (-t)+(-1,0)$) {};
\node[coordinate] (-a2) at ($ (-t)+(1,0)$) {};
}}}

\tikzset{pics/BS/.style args={#1}{code={

\node[station] (base) {#1};

\draw[line join=bevel] (base.100) -- (base.80) -- (base.110) -- (base.70) -- (base.north west) -- (base.north east);
\draw[line join=bevel] (base.100) -- (base.70) (base.110) -- (base.north east);

\draw[line cap=rect] ([yshift=0pt]base.north) [antenna=1];

}}}

\tikzset{pics/car/.style args={#1}{code={
    \node (-base) at (4.2,2){};
    \draw[very thick,#1, rounded corners=0.18ex,fill=black!20!blue!20!white,-]  (2.7,1.8) -- ++(1,0.7) -- ++(1.6,0) -- ++(0.6,-0.7) -- (2.7,1.8);
        \draw[thick,#1, fill=#1,thick,-] (4.2,2.5) -- (5.3,2.5)  -- ++(0.2,-0.23) -- (4.4,2.27) -- (4.4,1.8) --(4.2,1.8) -- cycle;
      \draw[thick,-]  (4.2,1.8) -- (4.2,2.5);
      \fill[draw=#1,fill=#1,rounded corners=0.6ex,very thick,-] (1.5,.5) -- ++(0,1) -- ++(1.2,0.3) --  ++(3,0) -- ++(1,0) -- ++(0,-1.3) -- (1.5,.5) -- cycle;
      \draw[draw=black,fill=gray!50,thick,-] (2.75,.5) circle (.5);
      \draw[draw=black,fill=gray!50,thick,-] (5.5,.5) circle (.5);
      \draw[draw=black,fill=gray!80,semithick,-] (2.75,.5) circle (.4);
      \draw[draw=black,fill=gray!80,semithick,-] (5.5,.5) circle (.4);
      \filldraw[draw=#1, fill=KITorange, semithick] (2.1,1.3) -- ++(150:.5) arc (155:205:0.5) -- cycle;
  
}}}

\tikzset{
  pobl/.style={
    inner sep=0pt, outer sep=0pt, fill=#1,
  },
  pobl gron/.style n args={2}{
    pobl=#1, rounded corners=#2,
  },
  pics/person/.style n args={3}{
    code={
      \node (-corff) [pobl=#1, minimum width=.25*#2, minimum height=.375*#2, rotate=#3, pic actions] {};
      \node (-pen) [minimum width=.3*#2, circle, pobl=#1, outer sep=.01*#2, anchor=south, rotate=#3, pic actions] at (-corff.north) {};
      \node (-coes dde) [pobl gron={#1}{1pt}, anchor=north west, minimum width=.12125*#2, minimum height=.25*#2, rotate=#3, pic actions] at (-corff.south west) {};
      \node [pobl=#1, anchor=north, minimum width=.12125*#2, minimum height=.15*#2, rotate=#3, pic actions] at (-coes dde.north) {};
      \node (-coes chwith) [pobl gron={#1}{1pt}, anchor=north east, minimum width=.12125*#2, minimum height=.25*#2, rotate=#3, pic actions] at (-corff.south east) {};
      \node [pobl=#1, anchor=north, minimum width=.12125*#2, minimum height=.15*#2, rotate=#3, pic actions] at (-coes chwith.north) {};
      \node (-braich dde) [pobl gron={#1}{.75pt}, minimum width=.075*#2, minimum height=.325*#2, outer sep=.0064*#2, anchor=north west, rotate=#3, pic actions] at (-corff.north east)  {};
      \node [pobl=#1, minimum width=.05*#2, minimum height=.2*#2, outer sep=.0064*#2, anchor=north west, rotate=#3, pic actions] at (-corff.north east) {};
      \node (-braich chwith) [pobl gron={#1}{.75pt}, minimum width=.075*#2, minimum height=.325*#2, outer sep=.0064*#2, anchor=north east, rotate=#3, pic actions] at (-corff.north west) {};
      \node [pobl=#1, minimum width=.0375*#2, minimum height=.2*#2, outer sep=.0064*#2, anchor=north east, rotate=#3, pic actions] at (-corff.north west) {};
      \node (-fit person) [fit={(-pen.north) (-braich dde.east) (-coes chwith.south) (-braich chwith.west)}] {};
      \node (-pwy) [below=25pt of -fit person, every pin] {\tikzpictext};
      \draw [every pin edge] (-fit person) -- (-pwy);
    };
  };
}

\tikzset{pics/phone/.style args={#1}{code={
    \node (-base) at (1,1.75) {};
      \draw[draw=black,fill=black!80,rounded corners=0.2ex,very thick] (0,0) -- ++(0,3.5) -- ++(2,0) --  ++(0,-3.5) -- ++(-2,0) -- cycle;
      \fill[thick, fill=KITblue] (0.2,0.2) -- ++(0,3.1)  -- ++(1.6,0) -- ++(0,-3.1) -- ++(-1.6,0) -- cycle;
      \fill[fill=#1] (0.65,2) rectangle (0.75,2.1);
      \fill[fill=#1] (0.85,2) rectangle (0.95,2.3);
      \fill[fill=#1] (1.05,2) rectangle (1.15,2.5);
      \fill[fill=#1] (1.25,2) rectangle (1.35,2.7);
  
}}}

\newcounter{nsect}

\usetikzlibrary{backgrounds}

\def\BibTeX{{\rm B\kern-.05em{\sc i\kern-.025em b}\kern-.08em
    T\kern-.1667em\lower.7ex\hbox{E}\kern-.125emX}}
\begin{document}
\begin{acronym}[TROLOLO]
  \acro{ACM}{auto correlation matrix}
  \acro{ADC}{analog to digital converter}
  \acro{AE}{autoencoder}
  \acro{ASK}{amplitude shift keying}
  \acro{AoA}{angle of arrival}
  \acro{AoD}{angle of departure}
  \acro{AWGN}{additive white Gaussian noise}
  \acro{BER}{bit error rate}
  \acro{BCE}{binary cross entropy}
  \acro{BMI}{bit-wise mutual information}
  \acro{BPTT}{backpropagation through time}
  \acro{BPSK}{binary phase shift keying}
  \acro{BP}{backpropagation}
  \acro{BSC}{binary symmetric channel}
  \acro{CAZAC}{constant amplitude zero autocorrelation waveform}
  \acro{CDF}{cumulative distribution function}
  \acro{CE}{cross entropy}
  \acro{CNN}{concolutional neural network}
  \acro{CP}{cyclic prefix}
  \acro{CRB}{Cramér-Rao bound}
  \acro{CRC}{cyclic redundancy check}
  \acro{CSI}{channel state information}
  \acro{CZT}{chirp Z-transform}
  \acro{DFT}{discrete Fourier transform}
  \acro{DNN}{deep neural network}
  \acro{DDD}{delay-Doppler domain}
  \acro{DoA}{degree of arrival}
  \acro{DOCSIS}{data over cable services}
  \acro{DPSK}{differential phase shift keying}
  \acro{DSL}{digital subscriber line}
  \acro{DSP}{digital signal processing}
  \acro{DTFT}{discrete-time Fourier transform}
  \acro{DVB}{digital video broadcasting}
  \acro{ELU}{exponential linear unit}
  \acro{EBM}{event-based measurement}
  \acro{ESPRIT}{Estimation of Signal Parameter via Rotational Invariance Techniques}
  \acro{FFNN}{feed-forward neural network}
  \acro{FFT}{fast Fourier transform}
  \acro{FIR}{finite impulse response}
  \acro{FLOP}{floating point operation}
  \acro{GD}{gradient descent}
  \acro{GRU}{gated recurrent unit}
  \acro{GF}{Galois field}
  \acro{GMM}{Gaussian mixture model}
  \acro{GMI}{generalized mutual information}
  \acro{ICI}{inter-channel interference}
  \acro{ICZT}{inverse chirp z transform}
  \acro{IDE}{integrated development environment}
  \acro{IDFT}{inverse discrete Fourier transform}
  \acro{IFFT}{inverse fast Fourier transform}
  \acro{IIR}{infinite impulse response}
  \acro{ISI}{inter-symbol interference}
  \acro{JCAS}{joint communication and sensing}
  \acro{KKT}{Karush-Kuhn-Tucker}
  \acro{kldiv}{Kullback-Leibler divergence}
  \acro{LDPC}{low-density parity-check}
  \acro{LLR}{log-likelihood ratio}
  \acro{LTE}{long-term evolution}
  \acro{LTI}{linear time-invariant}
  \acro{LR}{logistic regression}
  \acro{MAC}{multiply-accumulate}
  \acro{MAP}{maximum a posteriori}
  \acro{MLP}{multilayer perceptron}
  \acro{ML}{machine learning}
  \acro{MSE}{mean squared error}
  \acro{MLSE}{maximum-likelihood sequence estimation}
  \acro{MMSE}{miminum mean squared error}
  \acro{NN}{neural network}
  \acro{OFDM}{orthogonal frequency-division multiplexing}
  \acro{OLA}{overlap-add}
  \acro{PAPR}{peak-to-average-power ratio}
  \acro{PDF}{probability density function}
  \acro{pmf}{probability mass function}
  \acro{PSD}{power spectral density}
  \acro{PSK}{phase shift keying}
  \acro{QAM}{quadrature amplitude modulation}
  \acro{QPSK}{quadrature phase shift keying}
  \acro{radar}{radio detection and ranging}
  \acro{RC}{raised cosine}
  \acro{RCS}{radar cross section}
  \acro{RDM}{range-Doppler matrix}
  \acro{RMSE}{root mean squared error}
  \acro{RNN}{recurrent neural network}
  \acro{ROM}{read-only memory}
  \acro{RRC}{root raised cosine}
  \acro{RV}{random variable}
  \acro{SER}{symbol error rate}
  \acro{SNR}{signal-to-noise ratio}
  \acro{SINR}{signal-to-noise-and-interference ratio}
  \acro{SPA}{sum-product algorithm}
  \acro{UE}{user equipment}
  \acro{ULA}{uniform linear array}
  \acro{VCS}{version control system}
  \acro{WLAN}{wireless local area network}
  \acro{WSS}{wide-sense stationary}
  \acro{ZP}{zero-padding}
\end{acronym}

\title{
Improved Estimation Accuracy in OFDM-based\\ Joint Communication and Sensing through Kalman Tracking and Interpolation

\thanks{This work has received funding 
	in part from the European Research Council
	(ERC) under the European Union’s Horizon 2020 research and innovation
	programme (grant agreement No. 101001899) and in part 
	from the German
	Federal Ministry of Education and Research (BMBF) within the projects
	Open6GHub (grant agreement 16KISK010) and KOMSENS-6G (grant agreement 16KISK123).}

}

\author{\IEEEauthorblockN{Charlotte Muth, Leon Schmidt, Shrinivas Chimmalgi, and Laurent Schmalen\\}

\IEEEauthorblockA{Communications Engineering Lab (CEL), Karlsruhe Institute of Technology (KIT)\\ 
		Hertzstr. 16, 76187 Karlsruhe, Germany, 
		Email: \texttt{muth@kit.edu}\vspace*{-1ex}}
}

\maketitle

\begin{abstract}
We investigate a monostatic \ac{OFDM}-based \ac{JCAS} system for object tracking. Our setup consists of a transmitter and receiver equipped with an antenna array for fully digital beamforming.
The native resolution of range and velocity in all radar-like sensing, including \ac{OFDM} radar sensing, is limited by the observation time and bandwidth.
In this work, we improve the parameter estimates (estimates of range) through interpolation methods and tracking algorithms. We verify our method by comparing the \ac{RMSE} of the estimated range, velocity and angle and by comparing the mean Euclidean distance between the estimated and true position.
We demonstrate how both a Kalman filter for tracking, and interpolation methods using zero-padding and the \ac{CZT} improve the estimation error. We discuss the computational complexity of the different methods. We propose the KalmanCZT approach that combines tracking via Kalman filtering and interpolation via the \ac{CZT}, resulting in a solution with flexible resolution that significantly improves the range \ac{RMSE}.

\end{abstract}


\acresetall
\section{Introduction}
%%% Paragraph 1: Motivation
\Ac{JCAS} is a technology to be integrated into the new 6G standard \cite{Wild2021} that adds a sensor functionality to wireless communication devices. Different features of the environment can be sensed as transmit signals are reflected by passive objects resulting in multiple reflections of the same signal being received. These reflections can be identified and attributed to different objects in the environment, transforming devices for wireless communication into potential sensors. Different applications have been proposed including weather monitoring, network optimization, traffic assistance or drone detection.
Additionally, an improvement in energy and spectral efficiency is expected compared to deploying separate radar and communication systems.  

\Ac{OFDM} is especially suitable as a waveform for 6G \ac{JCAS} as it is a well established waveform for communication, as it enables robust communication in frequency-selective channels for low computational resources. Additionally, it has been studied for application in radar for over a decade~\cite{Braun2010}. However, precise radar estimates rely on sufficient observation time and a large signal bandwidth. Bandwidth remains a finite resource as different systems operate simultaneously and interference has to be prevented.
In 6G, we expect a limited bandwidth that is primarily chosen for communication, limiting the sensing resolution. In particular, the limited range resolution contributes severely to the estimation error when localizing objects that do not communicate.

If we want to capture a trajectory in time, limited resolution makes a single estimate inaccurate.
Methods such as tracking and interpolation offer more accurate estimates of the spatial environment. A well-designed interpolation approach improves the ability of the system to distinguish between closely spaced objects and increases the estimation precision, effectively increasing the detail of the sensed environment. Interpolation approaches for sensing include \ac{ZP}, the \ac{CZT}~\cite{Xu2023}, or super-resolution methods~\cite{Liu2020a}. Tracking additionally helps to maintain consistent observations of moving targets, allowing for the prediction of their future positions, which is vital for navigation, collision avoidance, and beamforming.
If positions follow a trajectory, we can use prediction to refine the estimates, as the positions along a trajectory are typically correlated.
Numerous studies applying Kalman filters to object tracking are part of the radar literature, e.g. in FMCW radar~\cite{Lipka2019}. Especially, angle of arrival tracking has been of interest, since it enables precise beam tracking~\cite{Burghal2019}. True position tracking can potentially further enhance energy-aware beam tracking. Additionally, object tracking is a basic task for \ac{OFDM}-based \ac{JCAS} systems and therefore of interest~\cite{Sanson2021} in order to track, e.g., obstacles in traffic or drones.
In order to enable real-time tracking, the complexity of the resolution enhancement algorithm needs to be sufficiently low, yet it should yield a position estimate that is as accurate as possible.

%%% Paragraph 3
%% state of the art

%%% Paragraph 4
%% specific topic that this work covers
This paper specifically compares Kalman filtering structures and interpolation methods to improve tracking estimation in an \ac{OFDM}-based \ac{JCAS} system. We discuss how the computational complexity can be limited and how the estimation results are affected by interpolation and tracking, respectively. We present an alternative event-based measurement approach and propose the KalmanCZT as a combination approach of interpolation and tracking.

%%% Paragraph 5: Roadmap paper
%
% The remainder of this paper is structured as follows: Section~\ref{sec:sysmodel} introduces the proposed \ac{JCAS} system model and Sec.~\ref{sec:methods} presents the different resolution enhancement methods including the proposed KalmanCZT approach. Section~\ref{sec:sims} evaluates the performance of the presented methods. 
% Lastly, in Sec.~\ref{sec:concl}, the conclusion of this work is presented.

% \emph{Notation:} $\mathbb{R}$ and $\mathbb{C}$ denote the sets of real and complex numbers, respectively. Sets are generally denoted as $\mathcal{X}$, with the magnitude of a set being $|\mathcal{X}|$. We denote vectors and matrices with boldface lowercase and uppercase letters, e.g. vector $\vect{x}$ and matrix $\mat{X}$. The single matrix element in the $n$th row and the $k$th column of matrix $\mat{X}$ is denoted as $x_{nk}$. The transpose and conjugate transpose of a matrix $\mat{X}$ is written as $\mat{X}^\top$ and $\mat{X}^H$, while the Hadamard product is indicated with operator $\odot$. The diagonal matrix $\mat{D}$ with diagonal entries $\vect{d}$ is denoted as $\text{diag}(\vect{d})$ and the all-one vector of length $N$ is denoted as $\boldsymbol{1}_N$.
% A complex normal distribution with mean $\mu$ and variance $\sigma^2$ is denoted as $\mathcal{CN}(\mu,\sigma^2)$. Random variables are denoted as $\mathsf{X}$, with mutual information $I(\mathsf{X}_1,\mathsf{X}_2)$, entropy $H(\mathsf{X})$ and cross entropy $H(\mathsf{X}_1 ||\mathsf{X}_2)$.

\section{System Model}\label{sec:sysmodel}
In this paper, we investigate a monostatic \ac{OFDM}-based \ac{JCAS} setup, where the transmitter and the sensing receiver are co-located, i.e., part of the same base station, and are equipped with \acp{ULA} to perform digital beamforming. Our goal is to track the \ac{AoA}, range and radial velocity of a passive object based on its reflection to track its location in a 2D plane. The transmit signal is simultaneously used to communicate with a \ac{UE} equipped with a single antenna in a \emph{separate} area of interest. We simulate a split transmit beam to enable communication, yet focus on the object tracking in this work. This approach enables us to use samples beyond the pilots for sensing which increases the accuracy for sensing. 
%The system block diagram is illustrated in Fig. \ref{fig:flowgraphtrain_mono}. 

\subsection{Transmitter}
We consider an \ac{OFDM} signal
with $N$ orthogonal sub-carriers with sub-carrier spacing $\Delta f = B/N = 1/(T_{\text{s}} N) = 1/T$, bandwidth $B$, sample duration $T_{\text{s}}$, and \ac{OFDM} symbol duration $T$. $M$ \ac{OFDM} symbol vectors are processed together in an \ac{OFDM} frame. The $NM$ transmit symbols of a frame are selected from the QPSK modulation alphabet and can be arranged as a matrix $\mat{X}$ of dimension $N\times M$. The notation $\mat{X}[n,m]$ denotes the element of $\mat{X}$ in the $n$th row and $m$th column, corresponding to the symbol of the $n$th sub-carrier of the $m$th OFDM symbol in a frame. A \ac{CP} of duration $T_{\text{cp}}$ is added to each symbol before transmitting the equivalent baseband signal~\cite{Braun2010}
% \todo[inline, color=yellow]{MH: I think I would omit the \mbox{($n = 1, \ldots, N, m = 1, \ldots, M$)} earlier or at least make it coherent with the following equation where you use start with m/n = 0 instead of m/n = 1}
\begin{equation*}
    s(t) = \frac{1}{\sqrt{N}} \sum_{m = 0}^{M-1} \sum_{n = 0}^{N-1} \mat{X}[n,m] \mathrm{e}^{\mathrm{j} 2\pi n \Delta f (t - m T_0)} g(t - mT_0),
\end{equation*}
where $T_0 = T_{\text{cp}}+T$ and $g(t)=1$ for $t \in [-T_{\text{cp}},T]$ and $0$ elsewhere is a rectangular pulse shaping filter.
We transmit the \ac{OFDM} signal at carrier frequency $f_{\text{c}}$.
The transmitter is equipped with $K$ antennas in a \ac{ULA} with antenna spacing $c_0/2f_{\text{c}}$, where $c_0$ represents the speed of light. In this work, we assume a genie-aided beamformer to isolate the performance of the receiver.
After beamforming with the beamforming function $\tilde{w}_a=\mathrm{e}^{-\j\pi a\sin\varphi}/2+\mathrm{e}^{-\j\pi a\sin\theta}/2$, with $\varphi$ being the \ac{AoD} of the tracked object, $\theta$ being the \ac{AoD} of the communication receiver, and antenna index $a\in[0,K\nobreak-1]$, the signal transmitted towards angle $\varphi$ can be expressed as
\begin{align}
    \vect{s}_{\varphi}(t) = \sum_{a=0}^{K-1} \tilde{w}_a s(t) \mathrm{e}^{\j\pi a\sin\varphi}.
\end{align}

\subsection{Channels}
A part of the transmit signal is received by the \ac{UE} while another part is reflected by the object of interest and reaches the sensing receiver co-located with the transmitter. The sensing channel is assumed to have one reflection caused by the tracked object with delay $\tau_{\mathrm{o}}$,\footnote{We do not distinguish between random variables and scalars notation-wise to maintain readability} associated Doppler shift $f_{\mathrm{D},{\mathrm{o}}}$ and factor $a_{\mathrm{o}} = 1$ representing the radar cross section, path loss and directivity gains. The received signal is given by
\begin{equation*}
    r(t) =  a_{\mathrm{o}} s(t - \tau_{\mathrm{o}}) \mathrm{e}^{ \j 2 \pi  f_{\mathrm{D},{\mathrm{o}}} t  }.
\end{equation*}
Finally the signal $r(t)$ is normalized to the power $\sigma^2_{s}$ and then received by the different antennas, resulting in
\begin{align}
    \vect{r}_{\varphi,a}(t) =  r(t) \mathrm{e}^{\j\pi a\sin\varphi}+ q(t),
\end{align}
with independent noise $q(t) \sim \mathcal{CN}(0,\sigma^2_{\text{ns}})$ at each antenna. 

\begin{figure}
    \centering
    \includegraphics{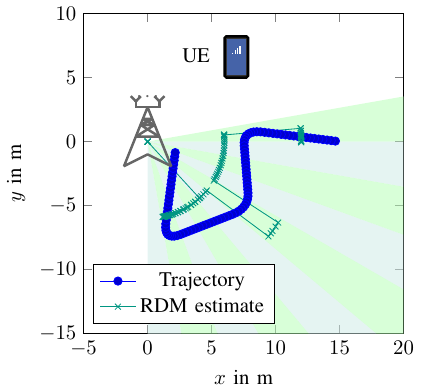}
    \caption{Scenario: An example trajectory of a passive object is shown with the corresponding position estimate using the native \ac{RDM}. The resolution of the \ac{RDM} is not able to accurately capture the object trajectory.}
    \label{fig:setup}
    \vspace{-5mm}
\end{figure}
\subsection{Sensing Receiver}
At the sensing receiver, we first estimate the angle $\varphi_{\text{est}}$ using the Bartlett method due to its computational complexity advantage~\cite{obeidat2022}.
Then, we perform receive beamforming based on $\varphi_{\text{est}}$. The received signal that is considered for estimating the radial velocity and the range, with $w_a=\mathrm{e}^{-\j\pi a\sin\varphi_{\text{est}}}$, is
\begin{align}
    \vect{r}_{\varphi_{\text{est}}}(t) =  \sum_{a=0}^{K-1} w_a(\varphi_{\text{est}}) \vect{r}_{\varphi,a} \mathrm{e}^{\j\pi a\sin\varphi}.
\end{align}
We can define the instantaneous \ac{SNR} for \ac{AoA} estimation to $\text{SNR}_{\varphi} = \sigma_{\text s}^2 / \sigma_{\text{ns}}^2$, with $\sigma_{\text{s}}^2$ representing the end-to-end received signal power. 

For estimating range and velocity, we follow the \ac{OFDM} sensing processing from~\cite{Braun2014}, i.e., we remove the cyclic prefix and transform the received signal to the time-frequency domain. After matched filtering, we obtain the received symbol matrix $\mat{R}$ and with the corresponding transmitted matrix $\mat{X}$, we can compute the element-wise ratio matrix $\mat{Y}=\mat{R}\oslash \mat{X}$, with $\oslash$ denoting the Hadamard division. The \acl{RDM} (RDM) is calculated as
\begin{equation}
    \mathrm{RDM} =  \underset{N \downarrow}{\mathrm{IFFT}} \left\{ \underset{M \rightarrow}{\mathrm{FFT}} \left\{\mat{Y} \right\} \right\},
    \label{eq:ch2:RDM}
\end{equation}
with $M \rightarrow$ denoting that the given operation is performed on each row of the input matrix and $N \downarrow$ denoting that the operation is performed on each column of the matrix.
We estimate the range and velocity as the values $\hat{r}$ and $\hat{v}$ corresponding to the element of the RDM with maximum power, since we assume that there is always one object present. This peak detection is used for all methods described in this work, a discussion of different detectors such as super-resolution methods is left for future work due to the difficulty in computing their exact computation complexity. In Fig.~\ref{fig:setup}, an example trajectory and the resulting estimated positions are shown. 

\section{Sensing Signal Processing}\label{sec:methods}
An \ac{OFDM}-based \ac{JCAS} system performing sensing on the received frame, as described in~\cite{Braun2014}, suffers from a limited resolution, as the native 2D grid of the \ac{OFDM} frame allows distinguishing between bins of spacing
\begin{align}
            r_{\text{res}} &= \frac{c_0}{2 B},\\
            v_{\text{res}} &= \frac{c_0}{2 f_{\mathrm{c}} MT_0}.
\end{align}
The complexity of computing the \ac{RDM} can be inferred from the \ac{FFT} complexity as $5NM\cdot\nobreak\log_2(NM)$ \acp{FLOP}~\cite[Table 2]{Chimmalgi2019}.
Signal estimation can be improved through interpolation methods, including \ac{ZP} of the frame or applying \acp{CZT}, or by using time correlation in the positions and applying tracking algorithms. In this section, we will introduce these methods in detail.

\subsection{Interpolation Methods}\label{sec:interpol}
\subsubsection{\Acl{ZP} (ZP)}
The application of \ac{ZP} before applying the \ac{IFFT} and the \ac{FFT} in the sensing processing \eqref{eq:ch2:RDM} is equivalent to a sinc-interpolation of the \ac{RDM}. Both the required memory for the transform result as well as the computational effort for calculating the interpolated \ac{RDM} increase with the \ac{ZP} amount. As we are mostly interested in improving the range estimation, we zero-pad the frame by adding rows to the input matrix resulting in a frame of dimensions $N_{\text{pad}} \times M$.
After \ac{ZP}, we need to calculate $N_{\text{pad}}$ \ac{IFFT} values from $N$ nonzero entries per original column, while the complexity of the \ac{FFT} per row doesn't change. 
Therefore, the total complexity can be approximated as $5MN_{\text{pad}}\log_2(N_{\text{pad}}) + 5NM\log_2(M)$ \acp{FLOP}.

\subsubsection{\Acl{CZT} (CZT)}
The \ac{CZT} generalizes the \ac{FFT} to frequency components beyond the unit circle~\cite{Sukhoy2019} and has been used to improve estimation in radar~\cite{Xu2023a}. In our application, we are interested in the zoom capabilities of the transform, i.e., we use it to analyze a smaller section of the \ac{RDM} in more detail. In the initial estimate, we use the whole \ac{RDM} as an observation window as we assume no knowledge of the initial position. In the subsequent steps, we use the last estimate as the center of the observation window.
The \ac{CZT} as well as its inverse can be calculated with a complexity of $75(2N\log_2(2N))$ \acp{FLOP}, assuming the number of output and input values is both $N$. Note that this is five times higher than the complexity reported for the \ac{CZT} in~\cite[Table 2]{Chimmalgi2019}. In~\cite{Chimmalgi2019}, the \ac{CZT} complexity is based only on the complexity of the underlying FFTs. However, we observe that the costs of other operations such as exponentiation are also significant. The total resulting complexity is $75M(2N\log_2(2N))+5NM\log_2(M)$ \acp{FLOP}.
%$\mathcal{O}(n\log{}n)$, while $n=\max\{N_{\text{FFT}},N_{\text{CZT}}\}$ with $N_{\text{FFT}}$ being the length of the original sequence and $N_{\text{CZT}}$ being the length of the transformed sequence~\cite{Sukhoy2019}.
The resulting transform to generate a zoomed in section is calculated using:
\begin{align}
%A &= i_{\text{est}}/N_{\text{FFT}}-d/2, \quad
%W = \mathrm{e}^{\left( \j \frac{2\pi d}{N_{\text{CZT}}}  \right)}\\
\mathrm{RDM}_{\text{fine}} &= \underset{[A_{\text{r}},W_{\text{r}},N] \downarrow }{\mathrm{CZT}} \left\{ \underset{[A_{\text{v}},W_{\text{v}},M] \rightarrow}{\mathrm{CZT}} \left\{\mat{Y} \right\}^* \right\}^*, \label{eq:cztsettings}
\end{align}
where the parameters are chosen as $W_{\text{r}} = \mathrm{e}^{\j2\pi r_{\text{CZT}} / ( N^2  r_{\text{res}})}$
$A_{\text{r}} = \mathrm{e}^{ \j2\pi \left(\frac{\hat{r}}{N r_{\text{res}}}-\frac{r_{\text{CZT}}}{2 N r_{\text{res}}}\right)}$, $W_{\text{v}}=\mathrm{e}^{\j2\pi/M}$ and $A_{\text{v}}=-1$, with the last range estimate $\hat{r}$ as the center of the range observation window of size $r_{\text{CZT}}$.
The range and velocity estimates corresponding to the element of $\mathrm{RDM}_{\text{fine}}$ with maximum power can then be obtained. These are more precise than estimation using the native \ac{RDM}.

%With the indices of $\mathrm{RDM}_{\text{fine}}$ with the highest power corresponding to the estimated range and velocity, we can calculate a higher precision estimate compared to the \ac{RDM}.
In our scenario, we use the \ac{CZT} to increase the range estimation accuracy.

\subsection{Tracking}
\subsubsection{Kalman Filter}
The Kalman filter can be divided into two distinct phases: the prediction phase and the update phase. In the prediction phase, the filter uses the current estimate $\vect{x}_{k-1}$ and the system dynamics $\mat{F}$ to predict the range, velocity and angle at the next time step. With our model, the prediction equation of the Kalman filter at time step $k$ is given by
\begin{align}
        \vect{x}_{k|k-1} =       \begin{pmatrix}  r_{k|k-1}\\ v_{k|k-1} \\ \varphi_{k|k-1} \end{pmatrix} =\underbrace{\begin{pmatrix}
            1 & -MT_0 & 0 \\
            0 & 1 & 0 \\
            0 & 0 & 1
        \end{pmatrix}}_{\mat{F}}
        \cdot \vect{x}_{k-1},\label{eq:kalman_pred}
\end{align}
with $\vect{x}_{k-1}$ denoting the output estimate of the Kalman filter in time step $k-1$ and $r_{k|k-1}$, $v_{k|k-1}$, $\varphi_{k|k-1}$ denoting the predicted range, radial velocity and \ac{AoA}, respectively.
As the radial velocity is defined to be positive for movements towards the receiver, we predict the new range as $r_{k|k-1}=r_{k-1}-MT_0v_{k-1}$.
During the update phase, the filter incorporates new measurements to correct the predicted state, weighting the impact of the measurement with Kalman gain $\mat{K}$. By combining these two phases, the Kalman filter continuously refines its estimates, leading to an increased accuracy. The behavior of the system is influenced by the observation noise $\mat{\Sigma}_{\text{meas},k}$, the noise caused by the resolution limit and the prediction noise $\mat{\Sigma}_{\text{pred},k}$. The computational complexity of a Kalman filter is
$\mathcal{O}_{\text{KF}} (s^3)$ with $s$ denoting the size of the state space (which is $3$ in our case).

\subsubsection{Event-Based Measurement (EBM)}
\acused{EBM}
% As we want a low-complexity solution to accurately track an object and the \ac{RDM} estimates $\hat{r}$ and $\hat{v}$ give us limited information, we propose an alternative estimation of the range that combines linear prediction with an event-based range measurement:
Seeking a low-complexity solution for accurate object tracking and recognizing that the \ac{RDM} estimates $\hat{r}$ and $\hat{v}$ provide limited information, we propose an alternative range estimation method that combines linear prediction with \aclp{EBM}:
\begin{align}
    r_{\text{ebm},k} = \begin{cases}
        \frac 12(\hat{r}_k + \hat{r}_{k-1}) &,\quad \hat{r}_k \neq\hat{r}_{k-1} \\
        r_{\text{ebm},k-1} - MT_0 \hat{v}_k &,\quad \hat{r}_k=\hat{r}_{k-1}
        % \only<4>{\\
        % r_{\text{meas}}[i-2] - T_{\text{est}} (\hat{v}[i]+\hat{v}[i-1]) &,\quad \hat{r}[i]\neq\hat{r}[i-1] \text{ and }\\ &\quad \hat{r}[i-1]\neq\hat{r}[i-2]}    
        \end{cases}
\end{align}
As the range resolution is low, we use the \ac{RDM} range estimate $\hat{r}_k$ if it changed compared to the last estimate $\hat{r}_{k-1}$ and set the output to the mean of $\hat{r}_k$ and $\hat{r}_{k-1}$. In all other cases, we rely on the linear prediction assuming a constant velocity of the target.
\begin{figure*}
    \centering
    \includegraphics{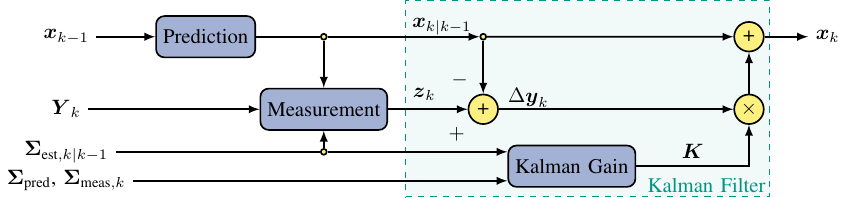}
    \vspace{-2mm}
    \caption{Flowgraph of the proposed joint tracking and filtering algorithm KalmanCZT}
    \label{fig:kalmanCZT-flowgraph}
    \vspace{-2mm}
\end{figure*}

\subsection{KalmanCZT}
We propose the KalmanCZT approach that combines tracking and filtering with a Kalman filter and the interpolation capabilities of the \ac{CZT}. We specifically address improving the range estimation of object tracking.
A high-level block diagram of the KalmanCZT is shown in Fig.~\ref{fig:kalmanCZT-flowgraph}. As the Kalman filter estimates the covariance matrix $\mat{\Sigma}_{\text{est},k|k-1}$ of the prediction, we use this estimate to determine the observation window of the \ac{CZT}, as well as using the predicted value $\vect{x}_{k|k-1}$ as the center of the observation window of the \ac{CZT}. 

In Algorithm~\ref{alg:kalmanCZT}, the KalmanCZT estimator is detailed. We first perform the prediction given by ~\eqref{eq:kalman_pred}, then feed the output values as the center of the observation window to the \ac{CZT}. The covariance matrix $\mat{\Sigma}_{\text{est},k|k-1}$ indicates the accuracy of the state estimates; therefore we use it to define the observation window of the \ac{CZT}.
As we want to ensure that the actual target range is in the observation window, we choose six times the standard deviation of the range as observation range $r_{\text{CZT}}$, using the three-sigmas rule. We add a minimum range of $1\,$cm. We calculate the \ac{CZT} parameters $W_{\text{r},k}$ according to this choice and $A_{\text{r},k}$ according to the predicted range-Doppler estimate $\vect{x}_{k|k-1}$. After applying the \ac{CZT}, we extract the range measurement $r_{\text{est}}$ and velocity measurement $v_{\text{est}}$ corresponding to the delay-Doppler observation with maximum power and combine them into an estimation vector $\vect{z}_k =\begin{pmatrix}
    r_{\text{est}} & v_{\text{est}} & \varphi_{\text{est}}
\end{pmatrix}^{\top}$.
As we assume the measurement noise to be mostly attributed to the resolution $d_{\text{CZT},k} = r_{\text{CZT},k} / N_{\text{CZT},1}$, we model it as uniform noise, resulting in the updated range variance
\begin{align*}
    \mat{\Sigma}_{\text{meas}}[0,0] = \int_{-d_{\text{CZT},k}/2}^{{d_{\text{CZT},k}/2}}  \frac{x^2}{d_{\text{CZT},k}} \dif x = \frac{d^2_{\text{CZT},k}}{12},
\end{align*}
while the other components of $\mat{\Sigma}_{\text{meas}}$ are not updated.
Finally, the Kalman gain $\mat{K}$ is updated based on the covariance matrices and the regular update step of the Kalman filter is performed.

\begin{algorithm}
\caption{KalmanCZT algorithm}\label{alg:kalmanCZT}
\begin{algorithmic}
\State Initialize state estimate $\vect{x}_{\text{track},0}$, covariance $\mat{\Sigma}_{\text{est},0|0}$ and prediction noise $\mat{\Sigma}_{\text{pred}}$
\For{each time step $k$}
    \State \textbf{Prediction step:}
    \State $\vect{x}_{k|k-1} = \mat{F} \vect{x}_{k-1}$
    \State $\mat{\Sigma}_{\text{est},k|k-1} = \mat{F} \mat{\Sigma}_{\text{est},k-1|k-1} \mat{F}^T + \mat{\Sigma}_{\text{pred}}$
    \State \textbf{CZT measurement:}
    \State $r_{\text{CZT},k} = \max \left(6 \sqrt{\mat{\Sigma}_{\text{est},k|k-1}[0,0]},1\,\text{cm} \right)$
    \State $d_{\text{CZT},k} = r_{\text{CZT},k} / N_{\text{CZT},1}$
    \State $W_{\text{r},k} = \mathrm{e}^{(\j2\pi r_{\text{CZT},k} / (N_{\text{CZT},1} N  r_{\text{res}}))}$
    \State $A_{\text{r},k} = \mathrm{e}^{ \j2\pi \max\left\{0,\frac{\vect{x}_{k|k-1}[0]}{N r_{\text{res}}}-\frac{r_{\text{CZT},k}}{2 N r_{\text{res}}}\right\}}$
    \State $\mat{Z}_k = \underset{[A_{\text{r},k},W_{\text{r},k},N_{\text{CZT},1}] \downarrow }{\mathrm{CZT}} \left\{ \underset{[A_{\text{v}},W_{\text{v}},N_{\text{CZT},2}] \rightarrow}{\mathrm{CZT}} \left\{\mat{R} \oslash \mat{X} \right\}^* \right\}^*$
    \State $r_{\text{est}}, v_{\text{est}}=$ Peak power detection $\left(\mat{Z}_k\right)$
    \State $\vect{z}_k = \left( r_{\text{est}}, v_{\text{est}}, \varphi_{\text{est}} \right)^{\top}$ 
    \State $\mat{\Sigma}_{\text{meas},k}[0,0] = \frac{d_{\text{CZT},k}^2}{12}$ 
    \State \textbf{Update step:}
    \State $\mat{K}_k = \mat{\Sigma}_{\text{est},k|k-1} ( \mat{\Sigma}_{\text{est},k|k-1} + \mat{\Sigma}_{\text{meas},k})^{-1}$
    \State $\vect{x}_k = \vect{x}_{k|k-1} + \mat{K}_k (\vect{z}_k -  \vect{x}_{k|k-1})$
    \State $\mat{\Sigma}_{\text{est},k|k} = (\mat{I} - \mat{K}_k ) \mat{\Sigma}_{\text{est},k|k-1}$
\EndFor
\end{algorithmic}
\end{algorithm}
\subsection{Performance Indicators}
We use the \ac{CRB} for benchmarking all sensing estimates. The \ac{CRB} is a fundamental lower bound of the estimation error, but does not consider time dependence of the parameters to be estimated. In a tracking scenario, we may get an estimation accuracy below the \ac{CRB}.
For the \ac{AoA}, the \ac{CRB} can be formulated according to~\cite[Ch. 8.4]{Trees2002} as
\begin{align}
\label{eq:CRB}
\text{Var}({\varphi_{\text{est}}}-\varphi) 
    %{C}_{\text{CR}} (\theta) 
    &= \frac{6\sigma_{\text{ns}}^2 \left(\sigma_{\text{ns}}^2 + K \sigma_{\text{s}^2} \right)}{\pi^2 \cos(\varphi)^2 N_{\text{win}} \sigma_{\text{s}}^4 K^2 (K^2-1)}.
\end{align}
We average over the full angle range using $\frac{1}{2\pi} \int_{-\pi}^{\pi} \cos(x)^2 \dif x = \frac 12$ to formulate an average \ac{CRB}
\begin{align}
\label{eq:CRBa}
    \overline{\text{Var}}(\varphi_{\text{est}}-\varphi) &= \frac{12\sigma_{\text{ns}}^2 \left(\sigma_{\text{ns}}^2 + K \sigma_{\text{s}}^2 \right)}{\pi^2 N_{\text{win}} \sigma_{\text{s}}^4 K^2 (K^2-1)}.
\end{align}
%For the angle estimation we use the Bartlett method of only the last $10\%$ of samples of the OFDM frame to limit calculation complexity.
For the range and velocity estimation bounds of the native \ac{RDM}, we bound the error to the quantization noise introduced by the resolution limit instead of the \ac{CRB}~\cite{Braun2011}. 

With the quantization introduced by the bins, the lower bounds are

\begin{align}
    \text{Var}(\hat{r}-r) &\geq \frac{c_0^2}{12(2N_{\text{FFT}}\Delta f)^2}\label{eq:bound-range}\quad \text{and}\\
    \text{Var}(\hat{v}_{\text{rad}}-v_{\text{rad}}) &\geq \frac{c_0^2}{12(2 f_c M_{\text{FFT}} T_0)^2}.
    \label{eq:crb-velocity}
\end{align}
%Since the achievable precision in the \ac{CZT} is equal to \ac{ZP} with the same bin range resolution, these results can be directly translated with  .

\section{Simulation Results}\label{sec:sims}
\subsection{Parameters}
We consider an \ac{OFDM} system with carrier frequency $f_\text{c}=5\,$GHz, a bandwidth of $B=25\,$MHz, and $N=2048$ subcarriers with a cyclic prefix length of $N_{\text{cp}}=30$. 
All simulation scenarios assume a constant $\mathrm{SNR}_{\varphi}=0\,$dB, independent of range variation.
We group $M=259$ OFDM symbols in one frame for sensing.
Therefore, the bounds \eqref{eq:CRBa}-\eqref{eq:crb-velocity} can be calculated as
\begin{align}
    \sqrt{\overline{\text{Var}}({\varphi_{\text{est}}}-\varphi)} &= 6.8 \cdot 10^{-4}\, (\text{rad}) \\
    \sqrt{{\text{Var}}(\hat{r}-r)} &= 1.73\, (\text{m})\\
    \sqrt{{\text{Var}}(\hat{v}_{\text{rad}}-{v}_{\text{rad}})} &= 0.409\, (\text{m/s}).%0.167\, (m/s)^2.
\end{align}
With the native \ac{RDM} resolution, the range is the most limiting factor to reconstruct the trajectory accurately while the \ac{AoA} can be estimated quite accurately, as illustrated in Fig.~\ref{fig:setup}.
%The achievable angle resolution is very good; yet the Bartlett method will not achieve this performance.
\subsection{Configuration of Interpolation Methods}
% We address the range resolution by choosing a smaller bin spacing at comparable complexities for \ac{ZP} and the \ac{CZT}. 
We choose $N_{\text{CZT}}=2048$ and use the complexity estimates in Tab.~\ref{tab:methods-comp-compl} to calculate the required \ac{FFT} length $N_{\text{pad}}=16N$ after \ac{ZP}, such that \ac{ZP} and the \ac{CZT} have a similar complexity.
%to achieve similar complexity of the \ac{ZP} to the \ac{CZT}.  

To get the same range resolution in the \ac{CZT} as in the \ac{ZP}, we configure the \ac{CZT} in~\eqref{eq:cztsettings} as
\begin{align*}
A_{\text{v}}&=-1, \quad W_{\text{v}}=\mathrm{e}^{\j2\pi/M}, \quad M=259 \quad \text{and} \\
A_{\text{r}}&=\mathrm{e}^{\j2\pi \left(\frac{\hat{r}}{Nr_{\text{res}}} - 2^{-5}\right) }, \quad W_{\text{r}}=\mathrm{e}^{\j\pi 2^{-14}}, \quad N_{\text{CZT}}=2^{11}.
\end{align*} 
%The parameter ${r}_{k-1}$ denotes the range estimated in the previous time step.
% A standard deviation in the range of $3.2\,$mm could potentially reached by superresolution methods, yet these estimators tend to be of quite high complexity. Enhancing the resolution to that amount would results increased memory and calculation efford for the \ac{FFT} or \ac{CZT}. 
% Without enhancing the resolution, we are bounded to a precision of
% \begin{align}
%     \bar{\text{Var}}(\hat{r}-r) &= 3\, (m)\\
%     \bar{\text{Var}}(\hat{v}_{\text{rad}}-{v}_{\text{rad}}) &= 0.167\, (m/s),
% \end{align}
% assuming a uniform distribution of positions.

\subsection{Configuration of Tracking Algorithms}
For the Kalman filter, the input values for the prediction and measurement covariance matrices significantly influence the behavior. Using experiments, we set them to
\begin{align}
    \mat{\Sigma}_{\text{meas}} &= \text{diag}\begin{pmatrix} 4.4 & 0.01 & 0.01 \end{pmatrix}^{\top}
    %\begin{pmatrix}
    %    4.4 & 0 & 0 \\
    %    0 & 0.01 & 0 \\
    %    0 & 0 & 0.01
    %\end{pmatrix}
    , \text{ and} \notag \\
    \mat{\Sigma}_{\text{pred}} &= \text{diag}\begin{pmatrix} 1.3\cdot 10^{-5} & 0.8 & 0.4 \end{pmatrix}^{\top}
    %\begin{pmatrix}
    %    1.3e{-5}& 0&  0 \\
    %    0& 0.8& 0 \\
    %    0& 0& 0.4
    %\end{pmatrix}
    ,
\end{align}
with $\text{diag}(\cdot)$ denoting the diagonal matrix operator.
We initialize $\mat{\Sigma}_{\text{est}}=\mat{0}$ if the initial position is known exactly and $\mat{\Sigma}_{\text{est}}=\text{diag}\begin{pmatrix}
    r_{\text{res}} & v_{\text{res}} & \frac 12
\end{pmatrix}^{\top}$ when using the native \ac{RDM} of the initial position is available.
For the KalmanCZT, we choose the same initial $\mat{\Sigma}_{\text{meas}}$, $\mat{\Sigma}_{\text{est}}$,  $\mat{\Sigma}_{\text{pred}}$ and prediction matrix $\mat{F}$.

% KalmanNet is trained on artificial data obtained by quantizing created trajectories to the resolution obtained by the \ac{RDM}. We choose an input size of the \ac{GRU} of $6$ and an output width of $9$. The results are reshaped to the intended dimensions by applying a fully connected layer before and after the \ac{GRU} with sigmoid activation.

% We use the truncated \ac{BPTT} with the last $30$ samples to train for the first half of training epochs and switch then to \ac{BPTT} for the full $92$ trajectory samples.
% %While training, we backpropagate after each estimate in order to keep the training overhead low. 
% In the initial training phases, we boost the estimation through using a genie-aided prediction, assuming that the tracked value in the last time step is the ground truth. After $25\%$ of training epochs, we disable the genie-aided tracking value. We train with a total of $10000$ trajectories with $92$ positions.

\subsection{Complexity Discussion}
We show a complexity estimate for each method in Tab.~\ref{tab:methods-comp-compl}. The \ac{RDM} has the lowest complexity, and we chose the parameters of \ac{ZP} and the \ac{CZT} to have similar complexity. On the other hand, the memory requirement for \ac{ZP} is $16$ times the memory required for the other approaches. The tracking algorithms introduce minimal additional complexity of $C_{\text{K}}\approx 50$ \acp{FLOP}.
%All algorithms are dominated by the complexity of the \ac{RDM} calculation. 
%Therefore we compare complexity relative to the \ac{RDM} complexity $C_0$. For least computational complexity, we get the \ac{CZT} and \ac{EBM}, yet the \ac{CZT} can give false results if we zoom into an area where the target is not present.
% Both tracking approaches have similar complexity, while the Kalman filter scales with the state space size, KalmanNet scales with the number of \ac{NN} nodes. The significant outlier in complexity is \ac{ZP}, needing $9C_0$ complex multiplications. 

\subsection{Performance Comparison}
\begin{table}
\caption{Complexity Comparison}
\centering
\label{tab:methods-comp-compl}
\begin{tabular}{lc}
\toprule
\multicolumn{1}{c}{Method}     & Complexity (\acp{FLOP})\\
\midrule
RDM    &    $C_0 = 5NM\log_2(NM)$     \\
Kalman filter     &      $C_0+C_{\text{K}}$      \\                    
EBM &  $C_0 + 3$  \\
ZP        &      $5MN_{\text{pad}}\log_2(N_{\text{pad}})+5NM\log_2(M)$          \\
CZT        &    $C_{\text{CZT}}=75M(2N\log_2(2N)) + 5NM\log_2(M)$ \\
KalmanCZT & $C_{\text{CZT}}+C_{\text{K}}$ \\
\bottomrule
\end{tabular}
\vspace{-0.2cm}
\end{table}

\begin{table}
\caption{\ac{RMSE} with known initial position}
\centering
\label{tab:methods-comp}
\begin{tabular}{lcccccc}
\toprule
\multicolumn{1}{c}{Method}     & range & velocity & \ac{AoA} & position &\\
\midrule
RDM                    &    $1.715$ m       &     \colorbox{KITgreen!30}{$0.40$ m/s}         &     \colorbox{KITgreen!30}{$0.08^{\circ}$}       &     $1.51$ m                              \\
Kalman filter                   &    $0.156$ m       &     \colorbox{KITgreen!30}{$0.40$ m/s}         &     \colorbox{KITgreen!30}{$0.08^{\circ}$}      &    $0.12$ m                               \\
EBM $\vphantom{\colorbox{KITgreen!30}{$0.333$ m}}$       &    $0.321$ m       &      \colorbox{KITgreen!30}{$0.40$ m/s}        &    \colorbox{KITgreen!30}{$0.08^{\circ}$}       &        $0.24$ m                           \\
ZP        &   $0.109$  m       &    \colorbox{KITgreen!30}{$0.40$ m/s}        &    \colorbox{KITgreen!30}{$0.08^{\circ}$}       &     $0.06$    m                          \\
CZT        &    $0.214$ m       &    \colorbox{KITgreen!30}{$0.40$ m/s}       &    \colorbox{KITgreen!30}{$0.08^{\circ}$}       &       $0.12$  m                           \\
KalmanCZT $\vphantom{\colorbox{KITgreen!30}{$0.15$ m}}$                      &     \colorbox{KITgreen!30}{$0.003$ m}      &    $0.41$ m/s          &   \colorbox{KITgreen!30}{$0.08^{\circ}$}        &             \colorbox{KITgreen!30}{$0.02$ m}                      \\ 
\bottomrule
\end{tabular}
\vspace{-5mm}
\end{table}

Assuming that the initial position is known exactly, as commonly assumed in the tracking literature, the \ac{RMSE} of the estimates for $1000$ trajectories consisting of $92$ points is shown in Tab.~\ref{tab:methods-comp}. In addition to the estimates, we show the mean Euclidean distance between estimated and true positions in the last column. The range and velocity error and of the \ac{RDM} and \ac{ZP} match the bounds of \eqref{eq:bound-range}, while the \ac{AoA} error does not reach the \ac{CRB}~\eqref{eq:CRBa}. This is expected behavior, since more intricate estimators such as super-resolution methods are needed to achieve performance close to \ac{CRB}.

For the range error, we achieve the lowest error of $0.003\,$m with the KalmanCZT, followed by \ac{ZP} with an \ac{RMSE} of $0.11\,$m and Kalman filtering with an \ac{RMSE} of $0.16\,$m.  
Notably, the \ac{EBM} also significantly improves performance down to an \ac{RMSE} of $0.32\,$m, while remaining the lowest complexity option.

We achieve a very similar performance for all velocity estimators. We have configured the \ac{CZT} and \ac{ZP} to result in the same velocity accuracy as the \ac{RDM}. As the velocity can fall off quickly, tracking methods do not improve estimation and in the case of KalmanCZT even slightly worsen it. This is most probably due to mismatched $\mat{\Sigma}_{\text{meas}}$ and $\mat{\Sigma}_{\text{pred}}$, and can be remedied by further tuning. Additional tracking does not seem to improve the \ac{AoA} estimate either. 

The position error is significantly reduced in all methods compared to the \ac{RDM} estimate of $1.51\,$m, with the  KalmanCZT resulting in the lowest error of $2\,$cm. A \ac{ZP} approach of similar complexity reaches an error of $6\,$cm, while all other approaches also allow for a lower mean Euclidean distance in a range of $12$-$24\,$cm. We get the smallest improvement by employing the \ac{EBM}.

\begin{figure}
\centering
\includegraphics{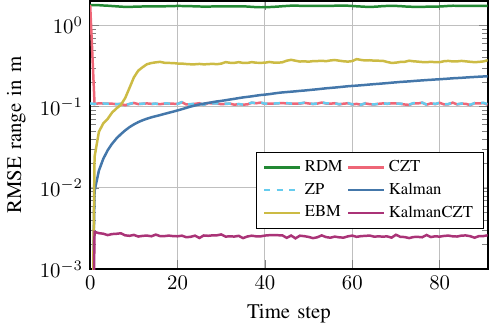}
\vspace{-0.2cm}
    \caption{Tracking accuracy over time RMSE vs. trajectory index with known start position}
    \label{fig:time-acc}
    \vspace{-0.5cm}
\end{figure}

In Fig.~\ref{fig:time-acc}, we show the \ac{RMSE} of the range estimate for different estimation and tracking variants averaged over $1000$ trajectories, under the assumption that the initial ground truth values are known. Therefore, the error in the first time step is zero for all algorithms with tracking component. The estimate based on the native \ac{RDM} performs worst and does not change over time. The Kalman filter starts with an error of zero that gradually increases up to $\approx$$0.25\,$m. The \ac{EBM} starts with a behavior similar to the Kalman filter but reaches its stable error plateau after $15$ samples and maintains an error of $0.33\,$m for the rest of the observation.
The performance of \ac{CZT} and \ac{ZP} is identical, except for the first time step where the \ac{CZT} performance equals the \ac{RDM} estimate, as it needs initialization of its observation window. The difference in the first time step is the reason for the higher \ac{RMSE} of \ac{CZT} in Tab.~\ref{tab:methods-comp} when compared to \ac{ZP}.
%Combining the \ac{EBM} with an interpolation method as the \ac{CZT}, we get an error slightly oscillating around the \ac{CZT} resolution. 
Lastly, leveraging not only a tracking mean but also a tracking variance in the KalmanCZT, the performance jumps from the initial estimate with an error of $0$ to an error plateau of $2.5\,$mm.

\begin{figure}
    \centering
    \includegraphics{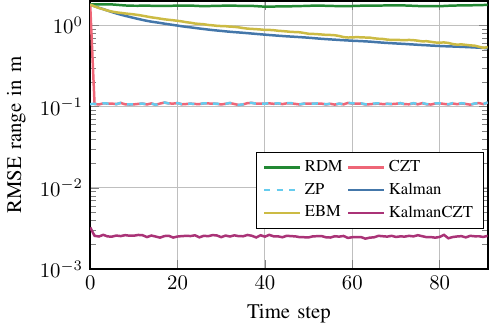}
\vspace{-0.2cm}
    \caption{Tracking accuracy over time RMSE vs trajectory index with start position being the \ac{RDM} estimate}
    \label{fig:time-acc-meas}
    \vspace{-0.5cm}
\end{figure}

Results for the case where the initial position is not fully known, but instead the initial \ac{RDM} estimate is used for all initial tracking estimates, are shown in Fig.~\ref{fig:time-acc-meas}. For the Kalman filter and the \ac{EBM}, we observe a decreasing error behavior from the initial \ac{RDM} accuracy in contrast to Fig.~\ref{fig:time-acc}. The \ac{EBM} shows a consistent small gap to the performance of the Kalman filter.
The KalmanCZT has the best performance, starting at at a slightly higher error but very quickly decreasing to the same error floor as in Fig.~\ref{fig:time-acc}. 
%Combining \ac{CZT} and \ac{EBM} can't improve the performance of the fine-grid measurement further.

\section{Conclusion}\label{sec:concl}
In this paper, we provide an analysis on estimation accuracy enhancing methods with a focus on range estimation in \ac{OFDM}-based \ac{JCAS} systems. We compare tracking methods and interpolation methods and propose the KalmanCZT approach, that combines estimation based on the \ac{CZT} with Kalman filtering. The range estimation error can be reduced from $\approx$$1.7\,$m to $\approx$$3\,$mm and the importance of tracking initialization is pointed out by comparing tracking with a known initial position and a known initial estimate.
We provide a detailed complexity analysis, noting that the additional complexity for tracking algorithms is very low compared to the computation of the \ac{RDM}. Application of the \ac{CZT} instead of the \ac{FFT} yields improved estimation, for a controlled increase in computational complexity. The KalmanCZT approach provides a flexible estimation controlled by the dynamics of the tracked object, enabling efficient use of the computational resources. 

For future investigations we are planning to extend the approach in order to apply it to real measurement data.
\acresetall
\acused{ESPRIT}

\bibliography{IEEEabrv,literature_short.bib}

\begin{thebibliography}{10}
\providecommand{\url}[1]{#1}
\csname url@samestyle\endcsname
\providecommand{\newblock}{\relax}
\providecommand{\bibinfo}[2]{#2}
\providecommand{\BIBentrySTDinterwordspacing}{\spaceskip=0pt\relax}
\providecommand{\BIBentryALTinterwordstretchfactor}{4}
\providecommand{\BIBentryALTinterwordspacing}{\spaceskip=\fontdimen2\font plus
\BIBentryALTinterwordstretchfactor\fontdimen3\font minus \fontdimen4\font\relax}
\providecommand{\BIBforeignlanguage}[2]{{%
\expandafter\ifx\csname l@#1\endcsname\relax
\typeout{** WARNING: IEEEtran.bst: No hyphenation pattern has been}%
\typeout{** loaded for the language `#1'. Using the pattern for}%
\typeout{** the default language instead.}%
\else
\language=\csname l@#1\endcsname
\fi
#2}}
\providecommand{\BIBdecl}{\relax}
\BIBdecl

\bibitem{Wild2021}
T.~Wild, V.~Braun, and H.~Viswanathan, ``Joint design of communication and sensing for beyond {5G} and {6G} systems,'' \emph{{IEEE} Access}, vol.~9, 2021.

\bibitem{Braun2010}
M.~Braun, C.~Sturm, and F.~K. Jondral, ``Maximum likelihood speed and distance estimation for {OFDM} radar,'' in \emph{Proc. {IEEE} Radar Conf.}, 2010.

\bibitem{Xu2023}
Y.~Xu, H.~Yi, W.~Zhang, and H.~Xu, ``An improved {CZT} algorithm for high-precision frequency estimation,'' \emph{Applied Sciences}, vol.~13, no.~3, p. 1907, Feb. 2023.

\bibitem{Liu2020a}
Y.~Liu, G.~Liao, Y.~Chen, J.~Xu, and Y.~Yin, ``Super-resolution range and velocity estimations with {OFDM} integrated radar and communications waveform,'' \emph{IEEE Trans. Veh. Technol.}, vol.~69, no.~10, Oct. 2020.

\bibitem{Lipka2019}
M.~Lipka, E.~Sippel, and M.~Vossiek, ``An extended {K}alman filter for direct, real-time, phase-based high precision indoor localization,'' \emph{IEEE Access}, vol.~7, 2019.

\bibitem{Burghal2019}
D.~Burghal, N.~A. Abbasi, and A.~F. Molisch, ``A machine learning solution for beam tracking in {mmWave} systems,'' in \emph{Proc. Asilomar Conf. Signals, Systems, and Computers}, Nov. 2019.

\bibitem{Sanson2021}
J.~B. Sanson, D.~Castanheira, A.~Gameiro, and P.~P. Monteiro, ``Cooperative method for distributed target tracking for {OFDM} radar with fusion of radar and communication information,'' \emph{IEEE Sens. J.}, vol.~21, no.~14, Jul. 2021.

\bibitem{obeidat2022}
H.~Obeidat, ``Performance comparisons of angle of arrival detection techniques using {ULA},'' \emph{Wireless Personal Communications}, vol. 126, no.~4, pp. 3611--3623, 2022.

\bibitem{Braun2014}
K.~M. Braun, ``\BIBforeignlanguage{english}{{OFDM} radar algorithms in mobile communication networks},'' Ph.D. dissertation, {Karlsruher Institut für Technologie (KIT)}, 2014.

\bibitem{Chimmalgi2019}
S.~Chimmalgi, P.~J. Prins, and S.~Wahls, ``Fast nonlinear {F}ourier transform algorithms using higher order exponential integrators,'' \emph{IEEE Access}, vol.~7, pp. 145\,161--145\,176, 2019.

\bibitem{Sukhoy2019}
V.~Sukhoy and A.~Stoytchev, ``Generalizing the inverse {FFT} off the unit circle,'' \emph{Scientific Reports}, vol.~9, no.~1, Oct. 2019.

\bibitem{Xu2023a}
Z.~Xu, S.~Qi, and P.~Zhang, ``A coherent {CZT}-based algorithm for high-accuracy ranging with {FMCW} radar,'' \emph{IEEE Trans. Instrum. Meas.}, vol.~72, 2023.

\bibitem{Trees2002}
H.~L. van Trees, \emph{{Optimum Array Processing: Part IV of Detection, Estimation, and Modulation Theory}}.\hskip 1em plus 0.5em minus 0.4em\relax Wiley, 2002.

\bibitem{Braun2011}
M.~Braun, C.~Sturm, and F.~K. Jondral, ``On the single-target accuracy of {OFDM} radar algorithms,'' in \emph{Proc. International Symposium on Personal, Indoor and Mobile Radio Communications ({PIMRC})}, 2011.

\end{thebibliography}
\bibliographystyle{IEEEtran}

\end{document}